# Ultrafast functional magnetic resonance imaging reveals neuroplasticity-driven timing modulations

Rita Gil, Francisca F. Fernandes, Noam Shemesh[*]

*Champalimaud Research, Champalimaud Centre for the Unknown, Lisbon, Portugal*

**\*Corresponding author:**
Dr. Noam Shemesh, Champalimaud Research, Champalimaud Centre for the Unknown
Av. Brasilia 1400-038, Lisbon, Portugal**.**
E-mail: noam.shemesh@neuro.fchampalimaud.org ;
Phone number: +351 210 480 000 ext. #4467.

**Running title:** Mapping activation sequence dynamics and their modulation upon dark rearing with ultrafast fMRI

**Word count:** 6228

**Author contributions:** RG stablished the dark rearing conditions; NS initiated and designed the ultrafast fMRI strategy; NS and RG established the required pulse sequences; RG performed all experiments and analysed the data; FFF provided support with experiments and analytical tools; NS and RG wrote the paper.

# Abstract

Functional Magnetic Resonance Imaging (fMRI) is predominantly harnessed for spatially mapping activation foci along distributed pathways. However, resolving dynamic information on activation sequence remains elusive. Here, we show an ultra-fast fMRI (ufMRI) approach – a facilitating non-invasive methodology for mapping Blood-Oxygenation-Level-Dependent (BOLD) response timings in distributed pathways with high spatiotemporal sensitivity and resolution. The mouse visual pathway was investigated under both normal and dark reared conditions. Results show that BOLD responses of normal reared mice preserve the neural input order from onset to peak times. However, modulatory effects in cortical responses due to dark rearing are only measurable at early response timings while a general delay in responses is measured at later timings. Our findings highlight the importance of robustly measuring early BOLD timings and pave the way for a better understanding and interpretation of functional MRI in healthy and aberrant conditions.

Neural activity underlying brain function is highly complex and spans several spatiotemporal orders of magnitude. Microscopic (sub)cellular interactions on the millisecond scale govern the neural code[1], while global scale millihertz oscillations between distributed neural masses underpin more advanced computations, behaviour, and cognition[2,3]. Whereas the former is typically studied invasively, via, e.g., electrophysiology and optical fluorescence imaging, the latter is mainly the domain of functional magnetic resonance imaging (fMRI)[4,5].

Taking advantage of the global view of the brain allowed by fMRI along with its non-invasive nature, this technique has been widely used to study and characterise brain functional deficits and neuroplasticity events after trauma or disease[6–8]. However, despite that fMRI provides invaluable spatial information on neural activity in the global brain, gaining insight into the activation dynamics in distributed neural pathways remains notoriously difficult, mainly due to the following four key issues: (1) neural pathways are spatially distributed, which necessitates the acquisition of multiple slices covering large volumes, which in turn limits the temporal resolution; (2) the ensuing BOLD responses are sparsely sampled, which highlight the activation plateau rather than the full signal dynamics where plasticity modulations might better represented; (3) although the magnitude of the peak BOLD signal may correlate with the LFP amplitudes[9,10], the BOLD dynamics can be dissociated from the underlying neural dynamics[11,12]; (4) the inherent low signal to noise, which requires still longer acquisition times to enable signal recovery. Nevertheless, a few studies have achieved relatively high temporal resolution with fMRI. Ogawa et al[13]. and Hirano et al.[14] were able to study impulse response properties using 310 and 250 milliseconds temporal resolution, respectively. Yu et al.[15] studied the differential contributions of cortical micro/macro-vasculature to the fMRI signal with a temporal resolution of 200 milliseconds. Silva and Koretsky[16] obtained an effective temporal resolution of 40 ms in a long fMRI experiment by permuting phase encoding and repetition loops, thereby enabling onset time mapping; earlier onsets in rat cortical layers IV and V were observed upon stimulation[16]. These experiments revealed that neural input order may be preserved in BOLD onset times. Later, Yu et al. pioneered "line scanning" fMRI[17] where, by eliminating one spatial dimension from the imaging sequence, a single line in the cortex could be scanned at a very high spatiotemporal resolution of 50 milliseconds and 50 μm, respectively. With such high temporal resolution, the onsets of BOLD fMRI cortical signals indeed coincided with the order of expected neural inputs to the different cortical layers in both somatosensory and barrel cortices, lending further credence to the hypothesis of closer neural tracking via BOLD onsets as compared with BOLD plateau or peak dynamics[17].

These earlier ultrafast fMRI studies focused on single cortical lines in rats, revealing the ensuing laminar activation dynamics. However, mapping the activation sequence in an entire distributed pathway has yet to be demonstrated, to our knowledge, mainly due to the stringent requirements on spatiotemporal resolution that cannot be addressed via single line scanning. To enable the investigation of activation sequence in more global pathways, we have used an ultrafast approach – ufMRI – (Figure 1), which rests on the following hypotheses: (1) acquiring a whole plane rather than only a single line does not "cost" heavily in terms of temporal resolution if carefully tailored single-shot MRI methods are used; (2) given that BOLD onset times are closer to neural activity than the full BOLD response[16–18], the activation order will be preserved even in distributed pathways. The ufMRI methodology requires a-priori knowledge of the spatial location of the pathway, which can be obtained from anatomy and/or a conventional fMRI measurement. Then, an oblique plane passing through the pathway's areas is designed, and, rather than acquiring the conventional numerous slices required to cover the pathway, only a single plane is acquired (Figure 1A). This can result in dramatic temporal gains, as the time can be used to image the same plane repeatedly (Figure 1B). To address the low sensitivity, we harness a state-of-the-art 4-element array cryoprobe, which facilitates sufficient signal-to-noise for observing activation dynamics even in a single animal with a single run (Figure 1C).

As a model pathway, we investigated the mouse visual pathway for activation dynamics at 9.4 T (Figure 1D) and looked at early times of the BOLD response such as onset and quarter-height times, and later timings such as half-height and peak times. The rodent visual system has been widely investigated and its activation sequence is very well described[19–21], making it an excellent candidate for ufMRI validation. After the light enters the mouse retina it is converted to electrical signals by photoreceptors and transmitted by the retinal ganglion cells (RGCs) crossing the midbrain at the optic chiasm. Direct retinal projections arrive both in the dorsal lateral geniculate nucleus of the thalamus (dLGN) and superior colliculus (SC) initiating two parallel pathways: the geniculate and extra-geniculate pathways, respectively (Figure 1D), which were here targeted for ufMRI. Projections from the dLGN arrive to V1 (mainly in layers IV, V and VI). In parallel, SC projects broadly to all cortical layers of both primary and higher order visual cortex areas indirectly through lateral posterior thalamic nucleus (LP).

The visual system normal development was perturbed by rearing animals in complete darkness until adulthood therefore eliminating any visual experience necessary for the critical period of plasticity[22,23] (around postnatal days 28 and 36[24]) to occur. Rodent dark rearing is known to induce several microstructural and functional changes along the entire visual pathway such as delayed optic nerve myelination[25,26], abnormal cell discharge patterns in LGN[25], decreased visual cortex thickness[25,26] and spine density at the apical dendrites of the L5 pyramidal cells[27,28], as well as compromised cortical excitatory/inhibitory[29,30] mechanisms leading to abnormal visual evoked potentials[30–32]. Yet, so far, due to spatial coverage limitations, most studies looked at individual structures, predominantly visual cortex[22,27–32], and not at the entire system. Additionally, regarding functional aberrations, most studies reported delayed and/or abnormal cortical visual evoked potentials[30–32].

With the ufMRI approach we can detect functional dynamics along the entire visual pathway and their modulations upon perturbation to the system, by measuring early BOLD responses time parameters considered to be closer to neural activity inputs in an area.

Our results reveal that (1) the two parallel pathways in the visual activation sequence could be inferred directly from ufMRI; (2) functional modulations upon dark rearing were detected along the entire pathway with broader and lower amplitude dark reared BOLD responses compared to normal reared group ones and; (3) different modulation dynamics were measured in early versus later BOLD response timings of the dark reared group, with the former showing shorter cortical delays approaching subcortical delays which, in their turn, were similar to the normal reared group ones, and the latter reporting an overall delayed response for all regions of interest with comparable delays in-between regions to normal reared responses.

Potential implications for future research and applications are discussed.

# Results

**ufMRI setup and visual pathway schematics.** The experiment timeline is depicted in Figure S1. The ufMRI approach requires an assessment of the pathways' spatial location. In this study, we have identified the involved areas anatomically (Figure S2) and then confirmed that these areas appeared active in conventional fMRI experiments. We then positioned the ufMRI imaging plane as shown in Figure 1A, capturing most of the geniculate and extrageniculate pathways (Figures 1D and S2), and performed conventional (slow) fMRI scans to ensure the imaging plane was properly position (data not shown).

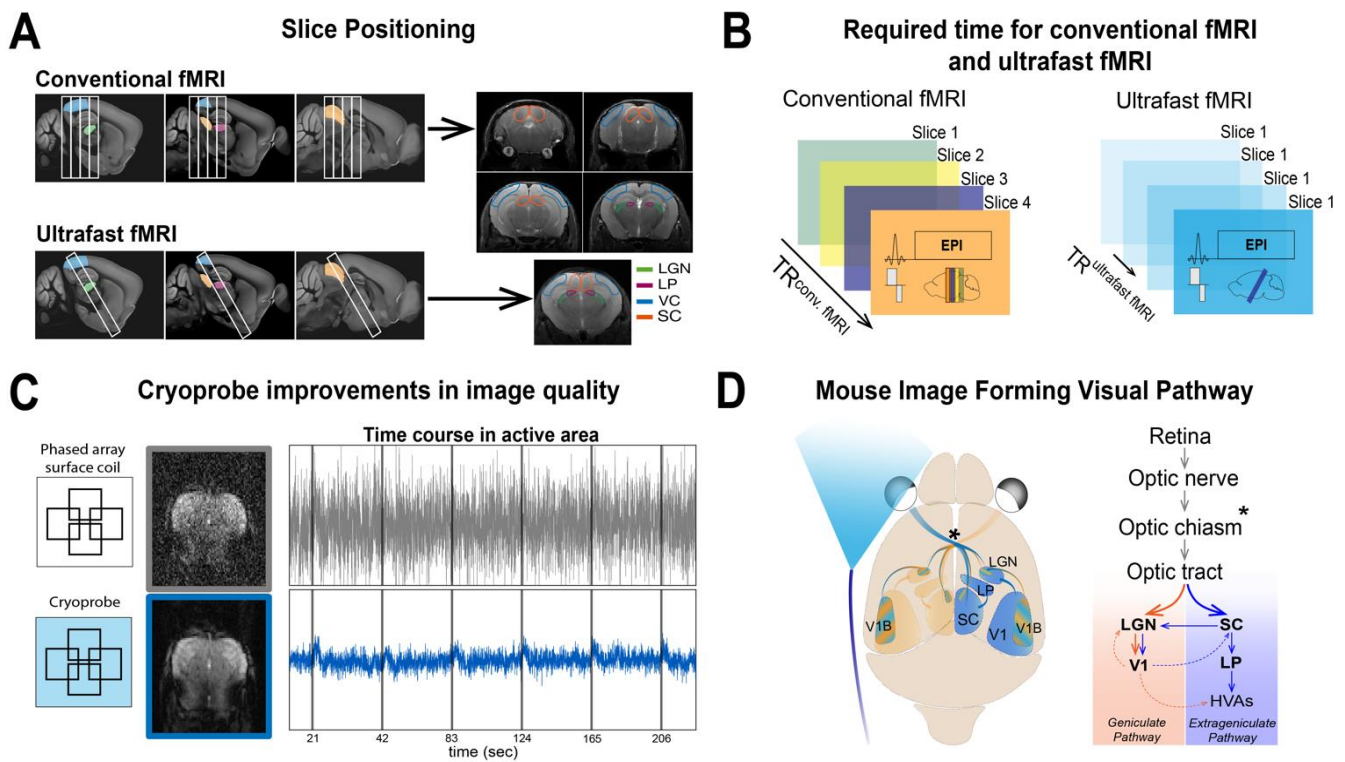

**Fig. 1 ufMRI concept and elements. (A)** Positioning of conventional fMRI and ufMRI slices. Conventional fMRI uses multiple slices for coverage of the important visual areas; by contrast, ufMRI captures all areas with a single oblique slice. **(B)** Timing of conventional fMRI vs. ufMRI approaches. The multiple slices required for coverage in conventional fMRI limit the temporal resolution while ufMRI can be repeated much more rapidly. **(C)** Signal to noise improvements due to the use of cryogenic reception coil. Top panel shows an ufMRI image degraded to room-temperature phased-array coil noise level and the ensuing temporal profile (gray), showing loss of functional signals. The bottom panel depicts the ufMRI image acquired in a (also phased-array) cryocoil, revealing robust activation signals (blue) along one run of the experiment. Vertical black bars represent stimulation. **(D)** Mouse visual pathway highlighting the involved cortical and subcortical areas upon monocular visual stimulation. Light enters the retina, followed by activity in two main pathways: the geniculate and the extrageniculate pathways. On the geniculate pathway, direct inputs to LGN project directly to V1, which then projects to higher visual cortical areas (HVAs). On the extrageniculate pathway, input arrives first to SC, which projects to the lateral posterior nucleus of the thalamus (LP) and to LGN. LP then projects to HVAs. The expected activation sequence is therefore LGN/SC -> LP -> V1 -> HVAs however, most HVAs are not expected to be active in flashing stimuli.

**Ultrafast functional MRI provides robust activation maps in the visual pathway for both rearing conditions.** We then investigated the ufMRI approach, first assessing the data quality and robustness. Figure S3 shows a movie of the raw ufMRI data in a single representative mouse, revealing no apparent motion or other artefacts. The signal to noise ratios (calculated from the normal reared mice group raw data) in contralateral V1, ipsilateral V1 (binocular region), LP, SC, and LGN were 73, 83, 47, 83 and 53, respectively. To avoid confounding effects in the general linear model (GLM) associated with the choice of the hemodynamic response function (HRF), the power in the fundamental frequency associated with the stimulation paradigm (Figure 2A) was mapped voxelwise (Figure 2B-C), revealing robust activation in all visual pathway junctions captured by ufMRI's slice in both rearing regimes. Areas not involved in the task did not exhibit task-driven activation.

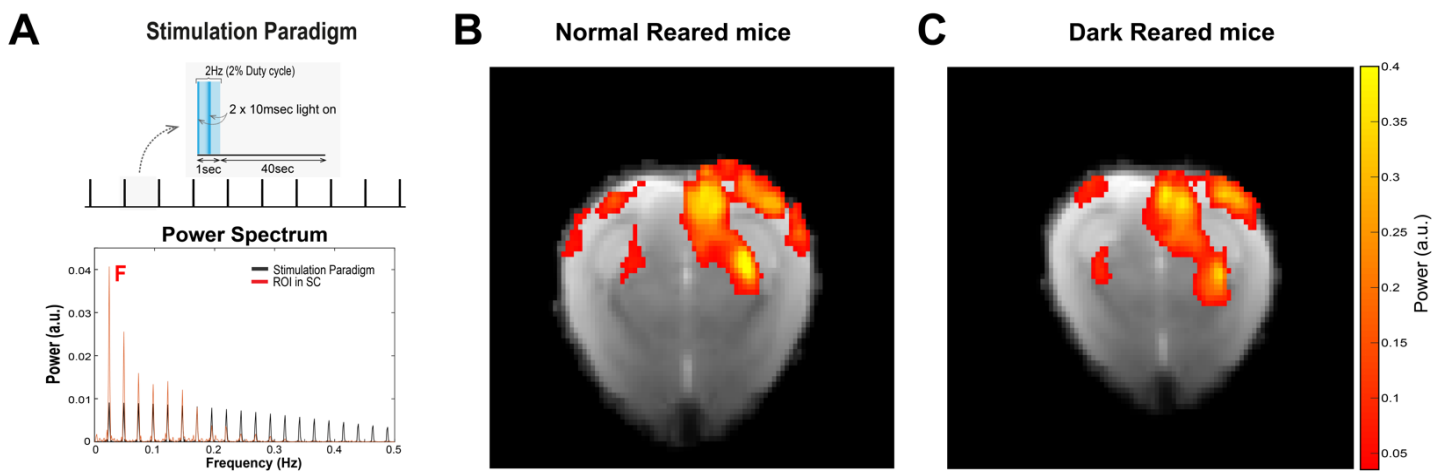

**Fig.2: Data-driven Fourier analysis. (A)** Top: Stimulation paradigm and building block. Bottom: Power spectrum of both the stimulation paradigm (black) and an ROI in SC of N = 5 normal reared mice (red). Clearly, there is a good agreement between the predicted and observed frequencies. To map the active areas, the fundamental frequency (marked in the plot of the power spectrum with the letter F in red) was selected and the integral was computed voxelwise and mapped. Corresponding power maps depicting robust activation in all areas of the visual pathway contralateral to the stimulation captured by the ufMRI slice in normal reared **(B)** and dark reared mice **(C)**.

**Inferring fast BOLD dynamics from ufMRI.** Once it was established that ufMRI provided robust BOLD responses, activation dynamics were assessed. Figure 3 shows BOLD dynamics in relevant ROIs for both normal (N=5) and dark (N=13) reared mice groups and responses in a single representative normal reared mouse are shown in Figure S5. Averaged ufMRI time courses for normal and dark reared mice groups are presented in Figure S4 and revealed robust activity in all contralateral ROIs associated with the geniculate and extrageniculate pathways with marked decreased amplitude in the dark reared group responses (Figure S4B). Remarkably, in the normal reared mice group, when looking at early BOLD response timings (up to half-height time) the following ROI sequence was observed: LGN and SC responded nearly together, followed by LP and then by contralateral and ipsilateral V1 (Figure 3A-C). Very similar dynamics were noted even in a single normal reared mouse, shown in Figure S5. Interestingly, dark reared group contralateral responses (Figure 3D-F) are broader than in the normal reared group and show a dramatic amplitude decrease. Moreover, the input sequence observed in the normal reared group early BOLD response timings is less clear after dark rearing. After normalizing each BOLD response to its maximum (Figure 3 C and F), two distinct dynamics appear in the normal reared mice group for the subcortical and cortical regions (Figure 3C) which is not present in the dark reared mice group (Figure 3F). Comparing each ROI separately (Figure 3G-O), the decrease in BOLD response amplitude for the dark reared group becomes clearer in all contralateral ROIs. Regarding the ipsilateral regions, BOLD responses appear to be similar in-between groups with a slight decrease in amplitude for the ipsilateral VC and the opposite effect observed for the ipsilateral SC. After normalizing individual ROI responses to their maximum (Figure S6), an increased response delay is seen in the dark rearing group for every contralateral ROIs (more pronounced in subcortical regions) while for ipsilateral ROIs similar responses exist for both groups.

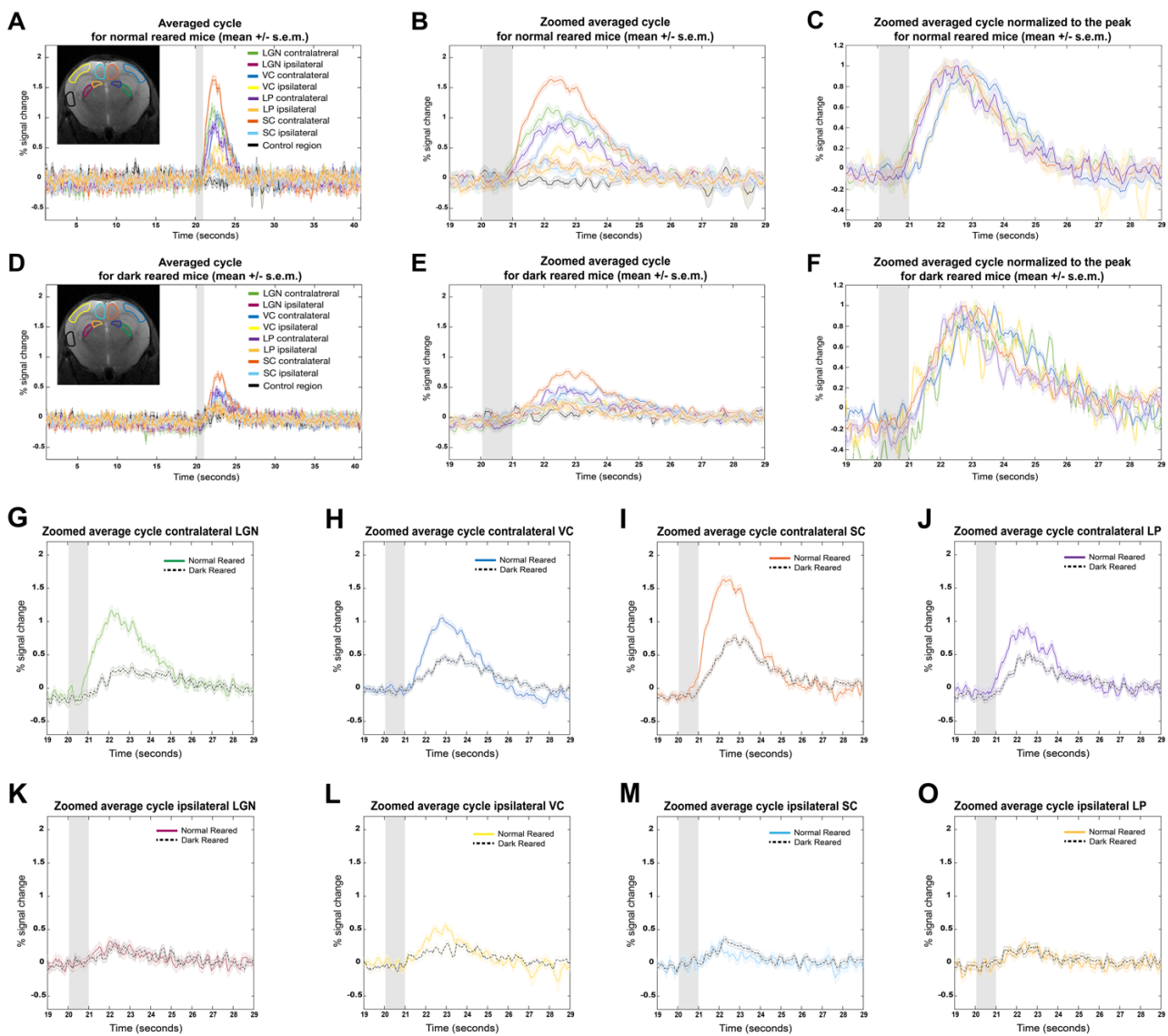

Fig.3: ROI Analysis for the normal (N = 5) and dark (N = 13) reared mice groups. Percentage signal change of the averaged BOLD response (mean +/- s.e.m.) for the different ROIs in the normal reared (A, B, C) and dark reared (D, E, F) mice groups. The anatomical image on the right top corner of (A) and (D), illustrates the different ROIs placement; Zoomed averaged BOLD response where different response profile times are more evident for the normal reared (B) and dark reared (E) mice groups; Normalized averaged response to the maximum percentage signal change of contralateral response profiles for the normal reared (C) and dark reared (E) mice groups. In the normal reared mice group, a clear distinction between cortical and subcortical normalized response timings can be observed and is absent in the dark reared group normalized responses. Grey bar represents the visual stimulus; Zoomed averaged BOLD responses for each individual ROI comparing the two different rearing conditions: contralateral and ipsilateral LGN (G, K), VC (H, L), SC (I, M) and LP (J, O). The decrease in BOLD amplitudes is visible for all contralateral responses while similar responses are observed in the ipsilateral hemisphere, with the exception of ipsilateral V1B where a BOLD amplitude decrease can be also seen.

**ufMRI enables robust quantification of BOLD timing parameters and reveals timing modulations upon dark rearing.**

Given the robustness of ufMRI data, the signal in every voxel was used to estimate timing maps such as onset, quarter-height, half-height and peak times (Figure 4). Figures 4A-D show the different timing maps for the normal reared mice group while Figures 4E-H shows similar maps for the dark reared group. The normal reared mice onset map (Figure 4A) shows earlier subcortical responses, nearly in parallel, around 500-700 ms and visual cortices activating only after 1000-1100 ms stimulus presentation. Note that, due to the slice obliqueness, cortical layers are traversed obliquely; still, the deeper (and more anterior) V1 layers reveal earlier onset times than the more ventral V1 layers. The onset map relative to the dark reared group (Figure 4E) shows similar delays between subcortical regions, between 500-900 ms, and cortical regions, between 700-1100 ms after stimulus presentation. As onset times are not trivial to define, a second early BOLD response timing was measured: the quarter to height time. Normal reared mice groups (Figure 4B) show subcortical quarter-height delays of 1100 ms while cortical delays are around 1600-1800 ms. On the other hand, dark rear group results (Figure 4F) show subcortical quarter-height delays of 1300-1600 ms and cortical delays above 1600 ms. At this time point an increased overall delay in the dark reared group responses is already observed, however, still with reduced differences between subcortical and cortical region delays when compared with normal reared group responses. At half-height time, the normal reared group (Figure 4C) exhibits timings around 1400 ms in subcortical areas and 1800-2100 ms for VC. When comparing these delays to the dark reared mice group ones (Figure 4G), an overall increased delay can be observed in all regions of the pathway: subcortical areas show timings around 1700-1900 ms while visual cortices show delays above 2100 ms. The separation of subcortical and cortical delays starts to become more evident at this time point. A similar general delay increase is visible in the peak maps between groups: Normal reared group delays (Figure 4D) present values below 2400 ms for subcortical regions and around 2800-3300 ms for VC while dark reared group (Figure 4H) subcortical delays increase up to around 2700-2900 ms and cortical delays to above 3000 ms.

The observations above are evidenced in the histograms of the different timings for the different groups (Figure 4I-Q). Note that, for the computations of the histograms, individual timing maps were used instead of the averaged maps shown in the two top rows of Figure 4. Onset time histograms relative to the normal reared group (Figure 4I), show subcortical regions with earlier onsets centred around 500 ms, while VC shows onsets times around 900-1000 ms. On the other hand, onset time histograms relative to the dark reared mice group (Figure 4N), show and overlap of timing delays for all ROIs, around 600 ms. Quarter-height time histograms (Figure 4K and 4P) show a slight general delay of responses in the dark reared group along with still a considerable decreased difference between cortical and subcortical delays when compared to normal reared responses. At half-height, the increased delays of cortical versus subcortical regions starts to be seen in the dark rear group histogram (Figure 4P) along with a more evidenced overall delay of dark reared responses compared to normal reared ones (Figure 4L). Regarding the normal reared group responses (Figure 4L), the difference between cortical and subcortical delays is maximum at half-height time. Finally, at the peak time (Figure 4M and 4Q), the dark reared responses delay is very clear and both groups present delayed cortical responses compared to subcortical ones.

**Vascular challenge to validate ufMRI-driven activation sequence.** To ensure that the differences observed in measured timing parameters within each group and in between groups are not merely a reflection of different vascular properties between locations and between different rearing conditions, N = 4 (13 runs averaged) normal reared mice and N=6 (5 runs averaged) dark reared mice were exposed to a hypercapnia challenge[33] during ufMRI experiments (Figure 5). For both rearing conditions, ufMRI signals exhibited nearly identical onsets for the different ROIs (Figures 5B and 5C) excluding the vasculature as the major contributing factor for the different measured timing parameters. Responses relative to the dark reared group are noisier than the ones relative to the normal reared group due to the fact that the former represent less than half of the number of averaged runs for the normal reared mice group.

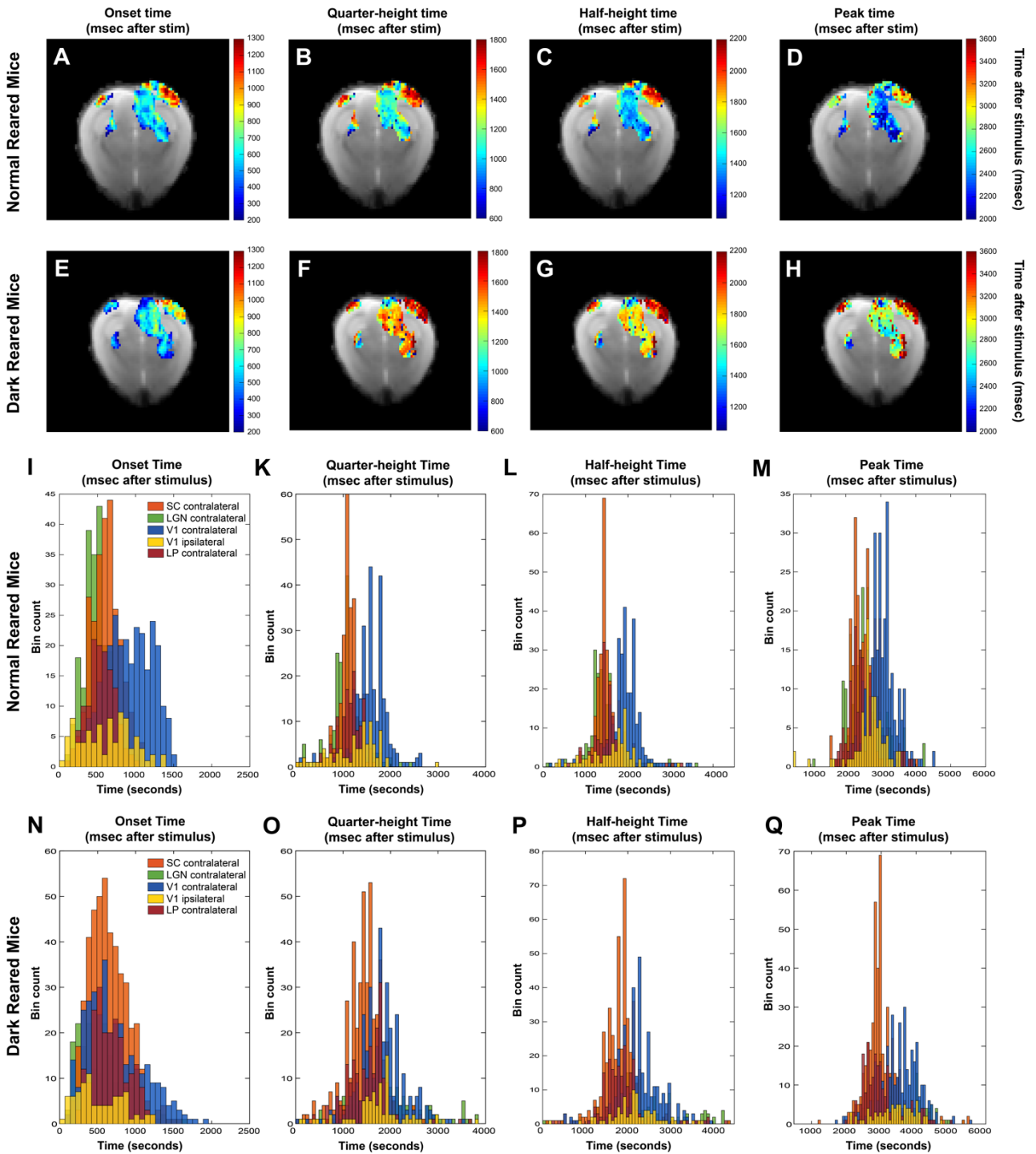

Fig.4: Voxelwise timing parameter analysis of the averaged cycles for normal reared (N=5) and dark reared (N=13) mice group. (A, E) Onset time maps; (B, F) Quarter-height time maps; (C, G) Half-height time maps and (D, H) Peak time maps for normal and dark reared groups respectively. The scale on the left side represents times relative to the beginning of the visual stimulus. Delayed times to quarter- height, half- height and peak of the BOLD responses can be observed in the dark reared group while onset time maps only show reduced timing differences between cortical and subcortical structures. Histograms for onset times (I, N); quarter- height times (K, O); half- height times (L, P) and peak times (M, Q) for normal and dark reared groups respectively. Histograms were computed from individual maps whereas the top shown maps are the averaged group maps. Onset time histograms reveal modulation of cortical timings relative subcortical ones upon dark rearing, with cortical onset timings approaching subcortical delays. The increased cortical delays seen in all measured timings of the normal reared mice, start to be more noticeable at the half- height time histogram for the dark reared mice (P) which highlights the importance measuring earlier BOLD timings. Peak time histograms between the two groups are the most similar ones but still with increased overall delays for the dark reared group.

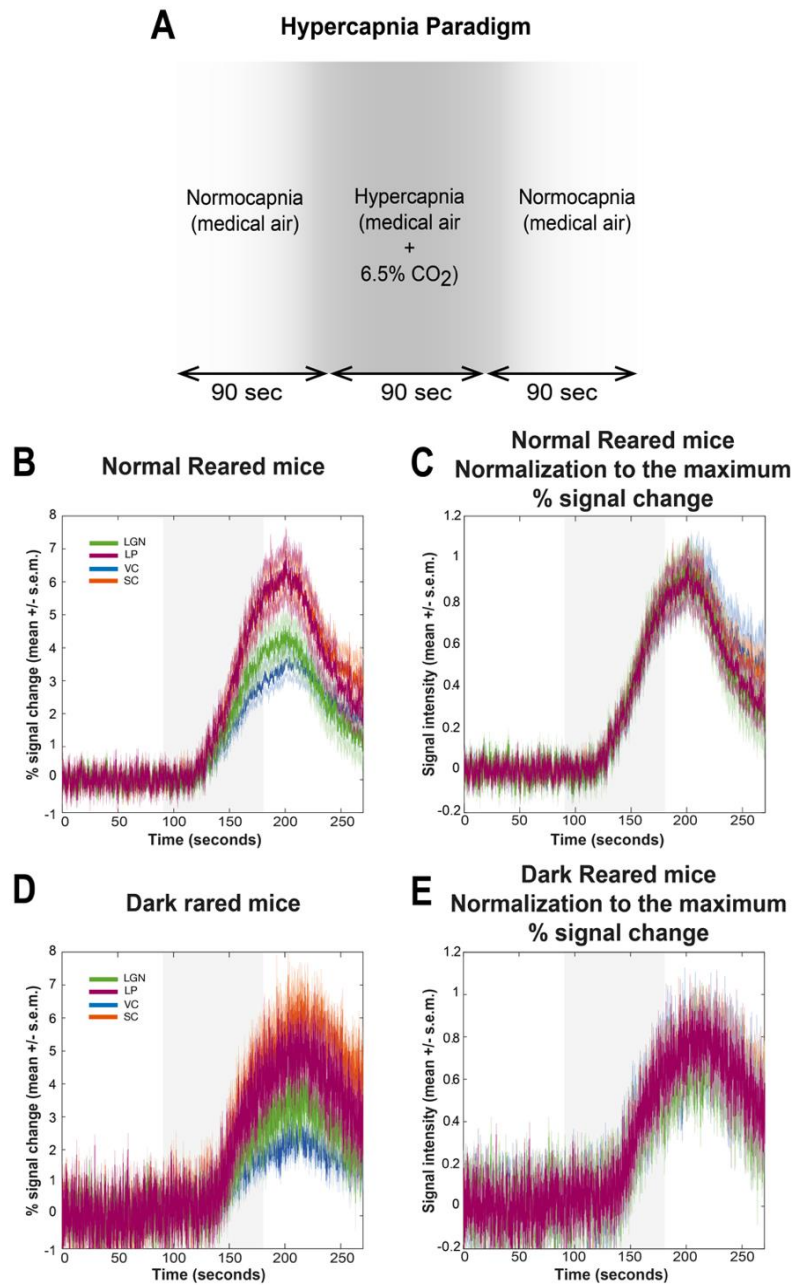

Figure 5: Hypercapnia experiment testing the dynamics of vascular responses. **(A)** Hypercapnia paradigm consisted of a manual switch, after 1.5 minutes of medical air, to a hypercapnic state with 6.5% $CO_2$ for 1.5 minutes. This was followed by a manual switch again to medical air for 1.5 minutes. Each run consisted in only one repetition of this block. Percentage signal change response profile (mean +/- s.e.m.) for different ROIs: VC, LGN, LP and SC of both hemispheres for the normal reared group **(B)** and the dark reared group **(D)**. Normalized response to the maximum percentage signal change for each ROI for the normal reared group **(C)** and the dark reared group **(E)**. Shaded areas indicate when the hypercapnic condition was applied. Note that the rise times are nearly identical for all areas, suggesting that the vascular response dynamics becomes dissociated only at later stages.

# Discussion

Contemporary functional MRI is mostly harnessed for spatially mapping activation foci rather than for deciphering dynamics between active regions mainly due to the inherent trade-off between spatial coverage and temporal resolution[5]. Prior studies have suggested that BOLD onset times accurately reflect the neural input order in cortical layers. However, the spatiotemporal resolution required for onset time mapping in distributed neural pathways has insofar eluded contemporary fMRI.

Here, we present ufMRI – an approach tailored for mapping the entire BOLD dynamics in distributed networks. The key element of ufMRI is the choice of an imaging plane passing through as many parts of the pathway as possible (e.g., acquiring a-priori either anatomically or conventional fMRI scans), thereby saving critical time typically lost in favour of more extensive spatial coverage. Coupled to ultrafast single-shot experiments resolving two-dimensional information and cryogenic coil sensitivity enhancements, we have shown that ufMRI achieves the temporal resolution and sensitivity required not just for resolving the early BOLD response times in distributed networks within the scanned plane, even in a single subject (Figures S5 and S6), but also to detect modulations of these timings upon perturbation of the normal functioning of the system by dark rearing the animals.

The mouse visual pathway was chosen as the first target for ufMRI as its anatomy and functional architecture are well established[19–21] and since perturbation of the system through dark rearing is easily to implement and minimally invasive to the animal. The visual system is wired in the following way: First SC and LGN receive the initial inputs from the retina, then, through the extrageniculate and geniculate pathways, LP receives inputs from SC , and V1 receives inputs from LGN[19–21] and LP.

For both rearing regimes, ufMRI reliably detected the active spatial locations in all visual areas (Figure 2) and, importantly, ufMRI-derived BOLD time dynamics for the normal reared mice group faithfully depicted what is expected from the neural inputs in the visual pathway: earlier times were measured for LGN and SC, followed by LP, and then V1 (Figure 3A-C). Additionally, the ufMRI approach proved sensitive enough so as to detect functional timing modulations upon dark rearing (Figures 3 and 4). It should be noticed that, in order to get comparable signal-to-noise ratio in each voxel's BOLD response for the two different groups, a larger number of animals was necessary for the dark reared group due to more sluggish responses.

The importance in mapping early BOLD timings, such as onset or quarter-height times, is illustrated in Figure 4. While later timing parameters, such as half-maximum or peak times, show a general delay in all BOLD responses upon dark rearing compared to normal reared BOLD responses, early timings reveal an additional modulation of cortical delays upon dark rearing that is lost at later timings. This is likely due to a more convoluted interplay between ongoing activity and hemodynamic timings at half-height and peak times which might hide interesting dynamics that are observed only at earlier timings. Further studies incorporating multimodal imaging[34–38] are required to elucidate the mechanisms underlying the differences between these early timings and later BOLD timing parameters. Still, early BOLD times mapping may benefit dynamic causal modelling or other model-based approaches for inferring dynamic information in fMRI.

One potential confounder of ufMRI is the vascular response dynamics: for example, if neural activity is identical in two regions, but vascular properties are dramatically different (e.g. due to variations in vessel density or reactivity), then the ensuing BOLD dynamics will vary between these areas. This limitation was directly tackled in this study by exposing the animals to a hypercapnia challenge typically used for calibrated fMRI approaches[33]. Hypercapnia induces strong vasodilation allowing for a relatively facile assessment of vascular component dynamics with little interference from the ongoing activity when applied moderately and for brief periods of time[39–41]. Our hypercapnia experiments revealed nearly identical rising timing profiles in the different areas under hypercapnia for the two animal groups (Figure 5). These findings are also consistent with autoradiographic measurements of CBF in the visual pathway areas of the normal reared mouse[42], which show very small differences in CBF (between 1.2-1.4 ml/100g/min in all areas). Taken together, our findings are inconsistent with different vascular responses as underlying sources for the differences observed in stimulus-induced BOLD early time maps.

What neural activity do ufMRI measured early BOLD times reflect? Neurovascular couplings are yet to have been fully deciphered[43–46] and the relationship between the BOLD responses and underlying neural activity remain contested[9,10,47–51]. Clearly, action potentials are orders of magnitude faster than BOLD and ufMRI does not map these directly. LFPs, representing synaptic activity and subthreshold processing, also occur on a faster timescale than the reported BOLD onset times: hundreds of milliseconds for BOLD onsets, versus tens of milliseconds for LFP onset delays recorded invasively under similar stimulation conditions[52]. Our stimulus included only two 10 ms flashes of light per event, separated by ~500 ms, suggesting that the delays are not due to (many) accumulating repeated inputs and vascular refractory periods[13]. Rather, it is conceivable that the neural inputs induce LFPs in each region at different rates, which can lead then to different energetic demands. These then trigger the (already slow) vascular responses at different times, thereby preserving the BOLD early delays. This mechanism is also consistent with the delays observed in the pioneering line-scanning fMRI experiments[17], which also revealed BOLD onsets only hundreds of milliseconds following (somatosensory) stimuli, and hundreds of milliseconds of delay in onset times between different cortical layers[16,17].

Our results show that BOLD responses in the visual pathway are affected by dark rearing.

Reported decreased and delayed optic nerve myelination, reduced number and diameter of RGCs[25,26], delayed and abnormal visual evoked potentials[30–32], along with delayed flash "off" responses[30] and reduced cortical vascular mass[25,26] are in line with the delayed and broad BOLD responses measured in the dark reared mice group. Additionally, the reduced BOLD response amplitudes upon dark rearing could be related with the already reported overall reduced responsiveness[30] of dark reared visual systems to flashing lights. Similar onset times of subcortical and cortical regions in the dark reared brain suggests a more immature and unspecialized system which is in line with reports of absence of ocular dominance and direction/orientation tuning in dark reared visual systems[22,30]. Future research incorporating multimodal approaches (e.g. ufMRI coupled with simultaneous electrical recordings[10], calcium recordings[37] and/or optical imaging[12,36]) will be required to further dissect the mechanisms underpinning BOLD time differences and modulations upon dark rearing.

Several limitations can be identified in ufMRI. First, many neural pathways may not be easily covered by a single plane, as required by ufMRI. In such cases, several more advanced solutions could be devised: for example, using multidimensional pulses[53] to excite 3D shapes corresponding only to the pathway's areas of interest. In addition, multiband pulses could be used to excite multiple planes at once[54], thereby offering a larger 3D coverage of the brain without any sacrifice for temporal resolution. Second, the fast acquisitions are strenuous on the gradient coils and amplifiers; in our study, we have actively monitored the amplifier temperature, and limited the temporal resolution based on the heating. It is interesting however to note that our study was therefore hardware-constrained rather than sensitivity-constraint and, for much shorter acquisition periods, much faster temporal resolution could be achieved with sufficient sensitivity for detecting even faster dynamics. This could be important for more directly detecting neural activity via MRI without relying on BOLD mechanisms. A third potential ufMRI limitation is specific-absorption-rate (SAR), which can effectively heat the imaged tissue. However, by imaging only a single plane, the number of radiofrequency pulses is actually comparable or even reduced compared with whole brain multislice acquisitions. In addition, ufMRI can benefit from low flip-angle acquisitions entailing lower SAR. Finally, we note that even the sensitivity enhancements offered by the cryogenic reception coil, which were extremely useful in this study, are not a prerequisite for ufMRI. In lieu of a cryocoil (or when the tissue noise is larger than coil noise, for example in human imaging), the acquisitions can be averaged if needed, with the sensitivity approximately growing as the square root of the number of averages. In addition, though we used the mouse as a model due to its relevance in biological research (e.g., due to advanced transgenesis), experiments in rats or other larger species should be easier to perform in terms of sensitivity as well as stability.

To conclude, here we show that the ufMRI approach is capable of robustly mapping activation sequence in distributed networks, and demonstrate its accuracy and utility for deciphering the activation dynamics in the mouse visual pathway in both normal and dark rearing. The spatiotemporal resolution offered by ufMRI was sufficient to map BOLD early times which, in normal rearing conditions, followed the correct neural input order but were modulated upon dark rearing of the animals. Hypercapnia experiments excluded differential vascular timings as the major source for ufMRI contrast. The generality of ufMRI suggests that it can play an important future role in understanding brain function in-vivo as well as for mapping aberrations in specific networks. These features augur well for future ufMRI applications.

# Acknowledgements

This study was funded in part by the European Research Council (ERC) (agreement No. 679058), as well as by Fundação para a Ciência e Tecnologia (Portugal), project 275-FCT-PTDC/BBB-IMG/5132/2014. The authors acknowledge the vivarium of the Champalimaud Centre for the Unknown, a facility of CONGENTO which is a research infrastructure co-financed by Lisboa Regional Operational Programme (Lisboa 2020), under the PORTUGAL 2020 Partnership Agreement through the European Regional Development Fund (ERDF) and Fundação para a Ciência e Tecnologia (Portugal), project LISBOA-01-0145-FEDER-022170. The authors would like to thank Dr. Daniel Nunes for assistance in the hypercapnia experiments.

## Methods

All animal care and experimental procedures were carried out according to the European Directive 2010/63 and pre-approved by the competent authorities, namely, the Champalimaud Animal Welfare Body and the Portuguese Direcção-Geral de Alimentação e Veterinária (DGAV). The experimental workflow is summarized in Figure S1: it consisted of an initial animal induction and beginning of sedation and preparation for imaging. Each fMRI session started with positioning of the animal and acquisition of anatomical reference scans which were followed by the novel ufMRI scans. Below we elaborate on each experimental phase, as well as on the subsequent data analysis.

*Animal Preparation*

This study used twenty-two C57BL/6 mice aged between 8 and 10 weeks old and weighing 20.8 ± 2.0 g with *ad libitum* access to food and water. Nine animals were raised in a 12h / 12h light/dark cycle while the remaining thirteen were born and reared in complete darkness until they reached adulthood.

Mice were prepared for anaesthesia and induced in a dedicated box allowing the flow of 5% isoflurane (Vetflurane, Virbac, France) for ~1.5 minutes and briefly maintained under a lighter isoflurane dosage of 2.5 - 3.5% during preparation for imaging. Two optic fibres connected to one blue LED ($\lambda$ = 470 nm and I = 8.1x10$^{-1}$ W/m$^2$) were placed near the left eye of the mouse while the right eye was covered by a custom 3D-printed eye patch (using black ABS-like resin from Formlabs, Massachusetts, USA) to achieve monocular visual stimulation.

Four to five minutes after induction, a bolus of medetomidine solution (1:10 dilution of 1 mg/ml medetomidine solution (VETPHARMA ANIMAL HEALTH S.L., Barcelona, Spain) - in saline) was administered by subcutaneous injection (bolus = 0.4 mg/Kg), and was, ten minutes later, followed by a constant infusion of 0.8 mg/kg/h, delivered via a syringe pump (GenieTouch, Kent Scientific, Torrington, Connecticut, USA). Before beginning a constant medetomidine infusion (14 – 16 min after induction), isoflurane dosage was progressively reduced and, at this point, was set to 0.25 - 0.35%, which was maintained constant throughout the remainder of the MRI session (Figure S1). During the entire time course of the experiments, animals were breathing oxygen-enriched medical air composed of 71% nitrogen, 28% oxygen and the remaining 1% comprise mostly argon, carbon dioxide and helium.

Throughout the experiments, the respiratory rate and temperature were monitored using a respiration pillow sensor (SA Instruments Inc., Stony Brook, USA) and an optic fiber rectal temperature probe (SA Instruments, Inc., Stony

Brook, New York, USA), respectively. Each experiment lasted around two and a half hours. In the end of the experiment, a 5 mg/ml solution of atipamezole hydrochloride (VETPHARMA ANIMAL HEALTH, S.L., Barcelona, Spain) was injected in the same volume as for the medetomidine bolus – 0.2 mg/kg – to revert the sedation. When scanning dark reared mice, all preparation steps were done with lights off with illumination coming only from two red LEDs placed above the animal's field of view.

Five normal reared mice underwent monocular visual stimulation while the other four underwent hypercapnia experiments. Regarding dark reared mice, thirteen animals underwent monocular visual stimulation and six of these animals also underwent hypercapnia experiments at the end of the MRI functional acquisitions.

*Visual stimulation setup*

An Arduino MEGA260 connected to the blue LEDs and receiving triggers from the MRI scanner, was used to generate square pulses of light. The optic fibres were placed horizontally in front of the left eye of the mouse at a distance of around 1 cm.

*Visual stimulation paradigm*

This study harnessed a block paradigm design starting with a dead-time period of 4 min and 21 seconds (a total of 5240 repetitions) while magnetization reached a steady state and gradient temperature was stabilized. This was followed by a resting period of 40 seconds and ten repetitions of a basic building block of the paradigm (Figure 2). The basic building block consisted of a visual stimulation epoch of 1 second at 2 Hz with 10 milliseconds pulse width (Figure 2, n.b. that the stimulation epoch only has two stimulation events in total during the 1 second "stim on" phase), followed by 40 seconds of rest. In total, a single paradigm was defined as [40 sec rest + [1 sec stim – 40 sec rest]$_{10}$], lasting 7 minutes and 30 seconds altogether. In between each visual stimulation run, the animals were allowed to rest for 7 min to avoid habituation.

*MRI acquisition*

All data in this study were acquired using a 9.4T Bruker BioSpin MRI scanner (Bruker, Karlsruhe,Germany) operating at a 1H frequency of 400.13 MHz and equipped with an AVANCE III HD console including a gradient unit capable of producing pulsed field gradients of up to 660 mT/m isotropically with a 120 µs rise time. Radiofrequency transmission was achieved using an 86 mm quadrature coil, while a 4-element array cryoprobe (Bruker, Fallanden, Switzerland) was used for reception. The software running on this scanner was ParaVision 6.0.1®.

**Positioning and pre-scans**

Following localizer scans ensuring the optimal positioning of the animal and routine adjustments for centre frequency, RF calibration, acquisition of B0 maps, and automatic shimming using the internal *MAPSHIM* routine, a high-definition anatomical T2-weighted Rapid Acquisition with Refocused Echoes (RARE) sequence (TR/TE = 2000/13.3 ms, RARE factor = 5, FOV = 20 × 16 mm², in-plane resolution = 80 × 80 µm², slice thickness = 500 µm, $t_{acq}$ = 1 min 18 sec) was acquired for accurate referencing. This was performed both with a coronal view and tilted view to ensure that the regions of interest (ROIs) were being captured in the same slice.

**Ultra-fast fMRI acquisitions**

A tailored oblique slice was prescribed such that SC, LP, LGN and V1 all passed through its plane (Figure 1 and S2). ufMRI was acquired with a gradient-echo EPI acquisition with a temporal resolution of 50 milliseconds. In particular, TR / TE = 50 / 17.5 msec, FOV = 16 × 12.35 mm, in-plane resolution = 167 × 167 µm², slice thickness = 1 mm. Nine thousand repetitions were acquired per run, leading to a total scan time of 7 min 30sec. Runs were repeated between two and four times for the five different mice.

**Hypercapnia Experiments**

To disentangle neuronal and vascular components in the observed response profiles, a hypercapnia experiment was performed where the acquisition parameters were identical as in the visual stimulation experiments. The "paradigm" however, consisted of 1.5 minutes ventilation with medical air followed by a manual switch to a hypercapnia state with 6.5% $CO_2$ for 1.5 minutes. In the end of the hypercapnia period, the $CO_2$ was switched off, the animals resumed breathing medical air and the data was acquired for 1.5 additional minutes (Figure 6A). In between hypercapnia "runs", the animals rested for 5 minutes. These experiments would elucidate on the vascular component of the response profile and rule out the possibility that the differences in the onset times between the different ROIs are purely of vascular origin.

**Data Analysis**

The high sensitivity endowed by the cryoprobe facilitated data-driven analysis in this study, thereby enabling us to avoid complicated assumption-based models. The data analysis included three main components: (1) Data-driven Fourier analysis which does not require an a-priori knowledge of the hemodynamic response function (HRF) for identifying active areas; (2) data-driven investigation of functional signals in a-priori defined anatomical ROIs placed in the different visual pathway components; (3) Voxelwise fits of the data to produce functional timing maps.

*Pre-processing*

Pre-processing steps included outlier detection (time points whose signal intensity was 4.5 times higher or lower than the standard deviation of the entire time course), which were corrected through interpolation from the entire time course; less than 0.5% of data points were interpolated. Motion correction was performed relative to the first repetition using Matlab's functions *imregister* and *imregtform*. The motion events were used as a nuisance regressors in a general linear model, thereby producing motion corrected images in which variance associated with motion was removed. Data were later denoised using a Total Generalized Variation regularizer with an alternating direction method of multipliers solver[55].

For the voxelwise analysis and data driven analysis, the ufMRI images arising from different animals were aligned to the same space so that they could later be averaged accurately. The alignment was performed by calculating the transform matrix that aligns the mean image of each run to the mean image of the first run acquired from an animal that was chosen as a reference, and applying this matrix to all repetitions from that run (since motion within each run had already been corrected previously). A 2D gaussian filter with a standard deviation of 0.7 and a kernel size of 5 was also applied to each repetition to enable a smoother summation of the images from the different animals.

*Data-driven Fourier analysis for detecting activation*

The periodicity of the paradigm allows for a relatively unbiased data-driven power spectrum analysis to detect the active brain areas, without having to assume HRFs. Therefore, the power spectrum of the stimulation paradigm was calculated (Figure 2A) and the fundamental frequency was identified. After wavelet denoising each voxel's time course with Matlab's *wdenoise* function, the power under this peak was calculated and mapped voxelwise (Figure 2B and C), resulting in paradigm-associated activation maps.

*ROI analysis*

For each animal, nine anatomical ROIs were selected for different visual pathway structures both contralateral and ipsilateral to stimulation side (SC, LP, LGN, V1 contralateral, V1B ipsilateral) as well as for a control region (around the secondary somatosensory cortex). These were drawn by hand using the 4th Edition of Paxinos & Franklin's mouse brain atlas[56] for guidance and can be seen in Figure 3 on the left top side of the A and D plots.

A 5th degree polynomial fit to the resting periods was used to remove low frequency trends for each run. The detrended data were then converted into percentage change and the ten individual cycles were separated and averaged across all runs, and then across all animals, providing the averaged response within each region. The averaged response was then smoothed using a Savitzky-Golay filter with a window size of 9 timepoints and polynomial order 3. The mean $\pm$ standard error of the mean for each ROI was calculated (Figure 3). In the normal reared group, out of 150 completed cycles, 2 were rejected due to excessive motion, which could not be corrected. In the dark reared group out of 400 cycles, 30 corresponding to 3 complete acquisitions where alignment of images was not successful, were excluded. The different averaged response profiles were also normalized to the maximum of each curve (Figure 3C and 3F).

*Voxelwise analysis*

We then aimed to generate maps reflecting timing parameters of the visually evoked BOLD responses (Figure 4). To achieve this, all cycles arising from each voxel were detrended and averaged over animals to compute the voxelwise averaged response. Images were then smoothed using Matlab's *imgaussfilt* function with a sigma of 0.7. The averaged individual cycle for each voxel was then wavelet denoised using Matlab's functions *wdenoise* and was then fitted to a 4-term gaussian function using Matlab's *fit* function ($R^2$ of the fit was conditioned to be above 0.7). For the calculation of the time to peak, the maximum of the fitted curve in an interval from the stimulation time until 4.5 seconds had passed was used but only if this value was higher than the noise level (calculated as the 75th percentile of the voxel time points before stimulation). Whenever the fitted gaussian presented more than one peak within the defined range (given that BOLD responses can sometimes exhibit plateaus or multiple peaks, c.f. Figure 3), a new corrected peak time was calculated as the time index corresponding to the 75th percentile of the fitted peak values (when calculated this way, the final peak will fall in between the fitted peak and will be more weighted towards the highest fitted peak). To calculate the quarter and half height times, the time to reach the closest data timepoint of quarter and half of the peak value was calculated (i.e., no interpolation was performed between 50 ms intervals representing the repetition time), respectively. The onset time was computed as the time corresponding to 7% increase of the fitted gaussian curve.

Individual time maps were also generated for the histogram computation (third and fourth panels of Figure 4). However, as noise levels were higher at the individual level, in the data-driven analysis the fundamental frequency along with the second and third harmonics were used to generate the individual activation maps.

For all voxelwise analyses, the activation maps generated by the data-driven Fourier analyses, whether for the group or individual analyses, were used as a mask to remove areas that were not identified as "active". Only clusters with a minimum of fifteen pixels were considered as "active".

*Hypercapnia data analysis*

Hypercapnia data analysis was performed similarly to the ROI analysis described previously. Several ROIs including right and left hemispheres of SC, LP, LGN and V1, were manually drawn for each animal. The time courses were detrended with a linear fit to the initial 1.5 min of the time course. The responses were averaged across animals, a Savitsky-Golay filter with a window size of 9 timepoints and polynomial order 3 was applied, and the mean $\pm$ standard error of the mean for each ROI was calculated (Figure 5 C and F). For an easier comparison of the different response profiles, the curves were normalized to the maximum of each curve (Figure 5 D and E).

Supplementary Figures for

# Ultrafast functional magnetic resonance imaging reveals neuroplasticity-driven timing modulations

Rita Gil, Francisca F. Fernandes and Noam Shemesh*

*Champalimaud Research, Champalimaud Centre for the Unknown, Lisbon, Portugal*

Experimental time course

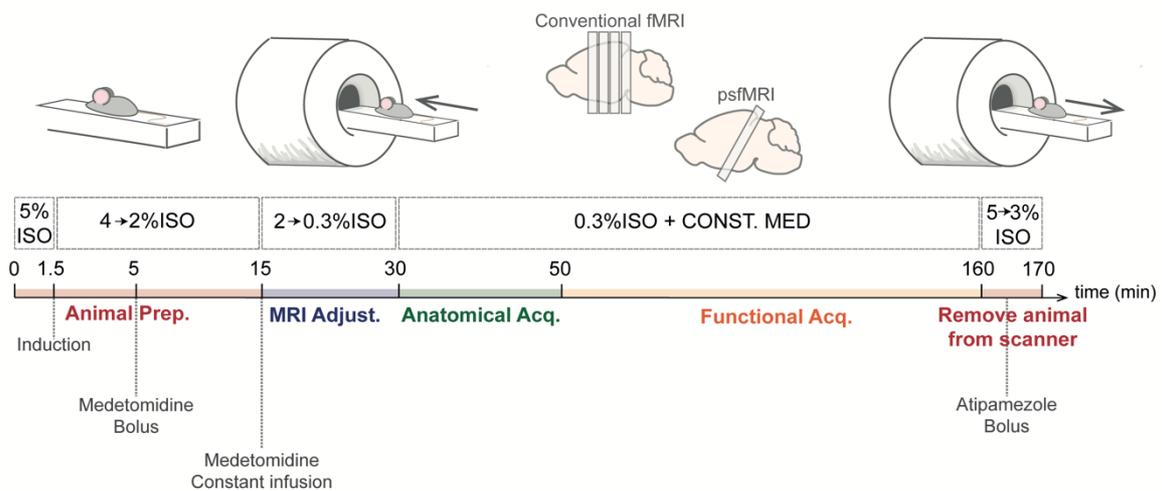

**Fig.S1: Experimental timeline.** The experiment commences with induction in a box breathing 5% isoflurane for ~90 seconds. This is followed by prepping the animal: head positioning is adjusted, transcutaneous injection of medetomidine is administered, rectal probe is inserted, and the right eye is covered by a 3D printed piece to facilitate monocular stimulation. The bolus of sedative is administered at 5-7 minutes after starting the induction and, the constant infusion starts ten minutes after, at around 15 minutes. At this point, the animal is ready and enters the MRI scanner. The first adjustments, which are accompanied by a reduction of isoflurane until the constant dosage of around 0.35% is reached, are performed. Anatomical scans scrutinizing optimal head positioning in the cryogenic coil, and acquisition slice placement are acquired until almost one hour has passed since the beginning of the induction. This allows for ample time for animal stabilization in the scanner. The functional scans are then acquired. After ~2.5 hours, the isoflurane dosage is increased again to 5% for safe removal of the animal; when the animal exits the scanner, the isoflurane dosage is reduced to 3% and an anti-sedative is administered, followed by full recovery in the animal's cage.

Regions included in the ufMRI tilted slice

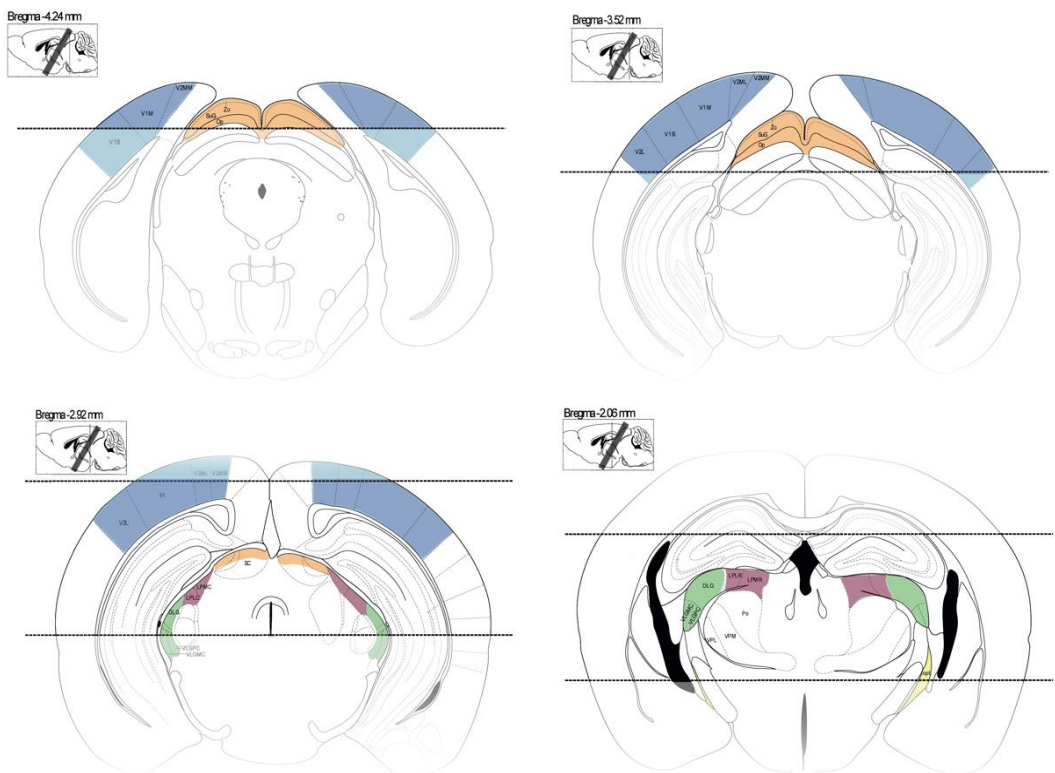

**Fig.S2: Detailed regions included in the oblique plane scanning slice.** Atlas slices with the detailed structures captured by the oblique psfMRI slice. Dashed lines mark the area included in the acquisition slice. Shown with different colours are the several areas captured by the oblique slice. The two top images show the posterior portion of the slice and the two bottom images show the most anterior portion of the slice.

GIF of one individual ufMRI run

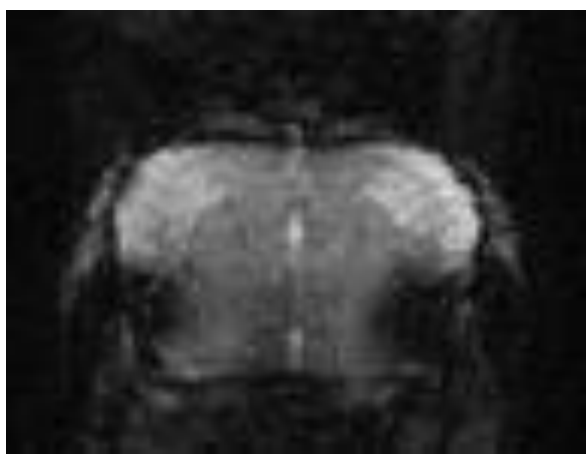

**Fig.S3: Raw data movie for one ufMRI run** in a representative animal showing quality of raw data. Very little motion or other artifacts were noted in these images.

Averaged time courses for different ROIs

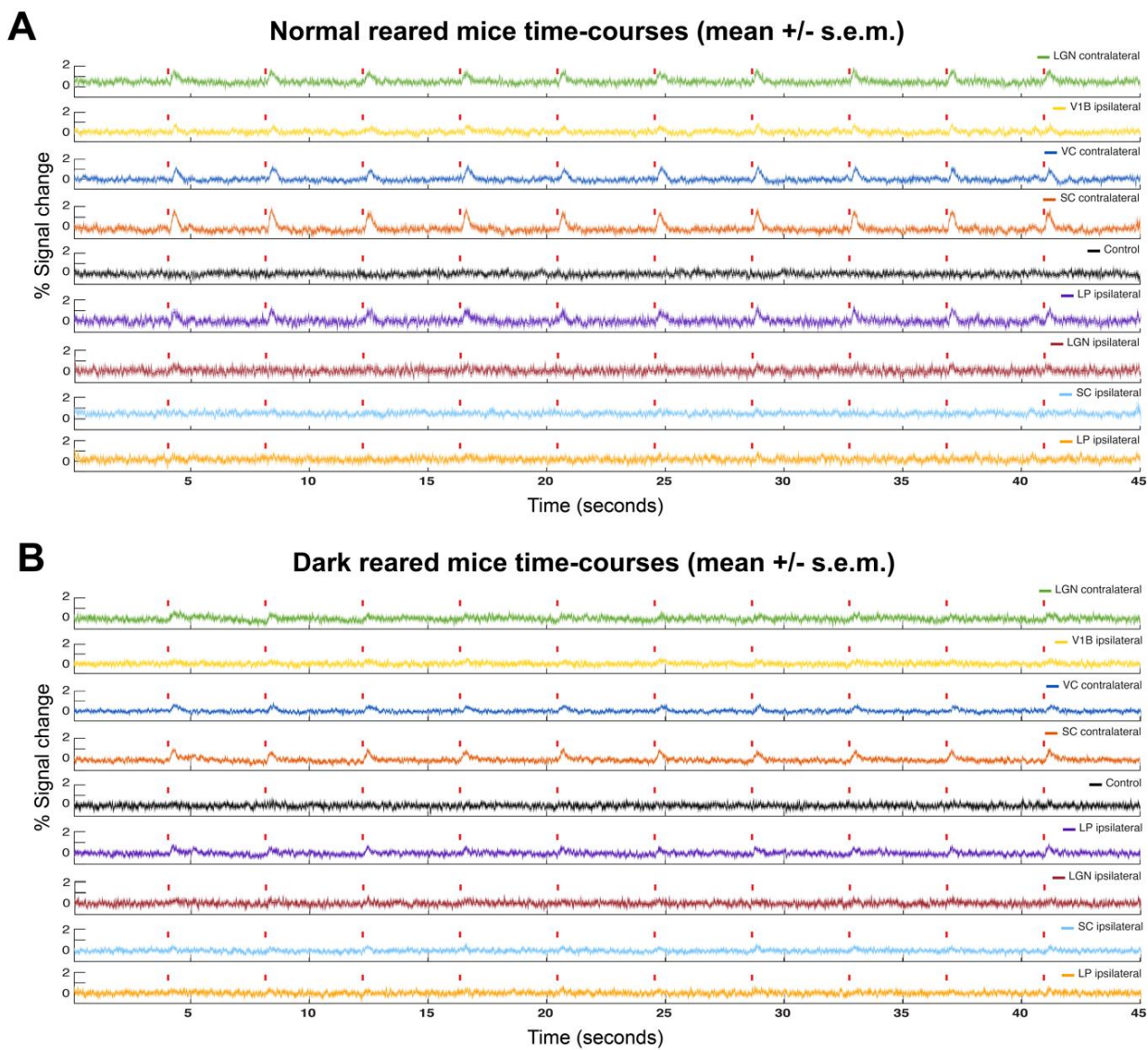

**Fig.S4: Average time courses (mean +/- s.e.m.):** Average time courses for different ROIs for normal **(A)** and dark reared **(B)** mice groups. Colour code: Green – LGN; Yellow – ipsilateral V1B; Dark blue: contralateral V1; Orange: SC; Black: Control region; Purple: contralateral LP; Dark red: ipsilateral LGN; Light blue: ipsilateral SC and light orange: ipsilateral LP. An overall decreased in the BOLD response amplitudes can be observed in all ROIs for the dark reared group. Red vertical marks represent individual visual stimulations which sum up to a total of ten stimulation blocks per time course.

Averaged time courses and averaged cycles for one individual normal reared mouse

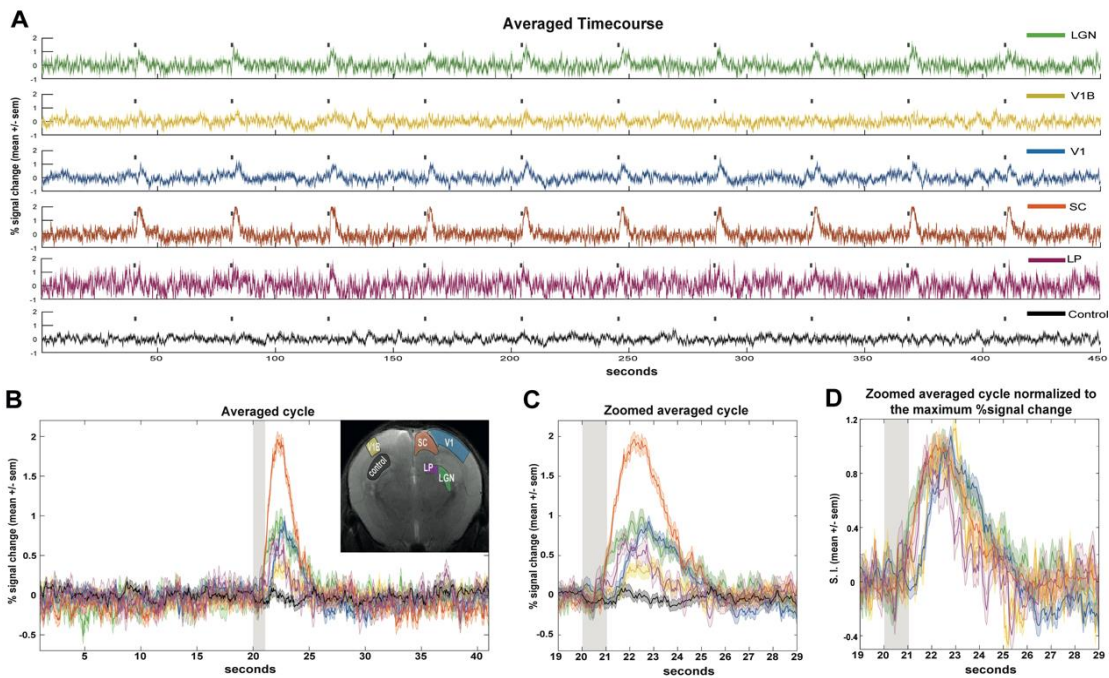

Fig.S5: ROI Analysis in a single normal reared animal (N = 1). **(A)** Percentage signal change for the averaged time course (mean +/- s.e.m.) consisting of ten stimulation blocks for the different ROIs: SC, LP, LGN, V1B of the ipsilateral hemisphere, V1 of the contralateral hemisphere and a control region on the ipsilateral hippocampus. Grey marks represent the visual stimulus. **(B)** Percentage signal change of the averaged BOLD response (mean +/- s.e.m.) for the different ROIs. The anatomical image on the right top corner shows the different ROI placement. **(C)** Zoomed averaged BOLD response where different response profile times are more evident. **(D)** Normalized averaged response to the maximum percentage signal change of each response profile. Grey bar represents the visual stimulus. ufMRI is sufficiently sensitive to detect the activation even in a single subject.

Individual ROI averaged responses normalized to their maximum

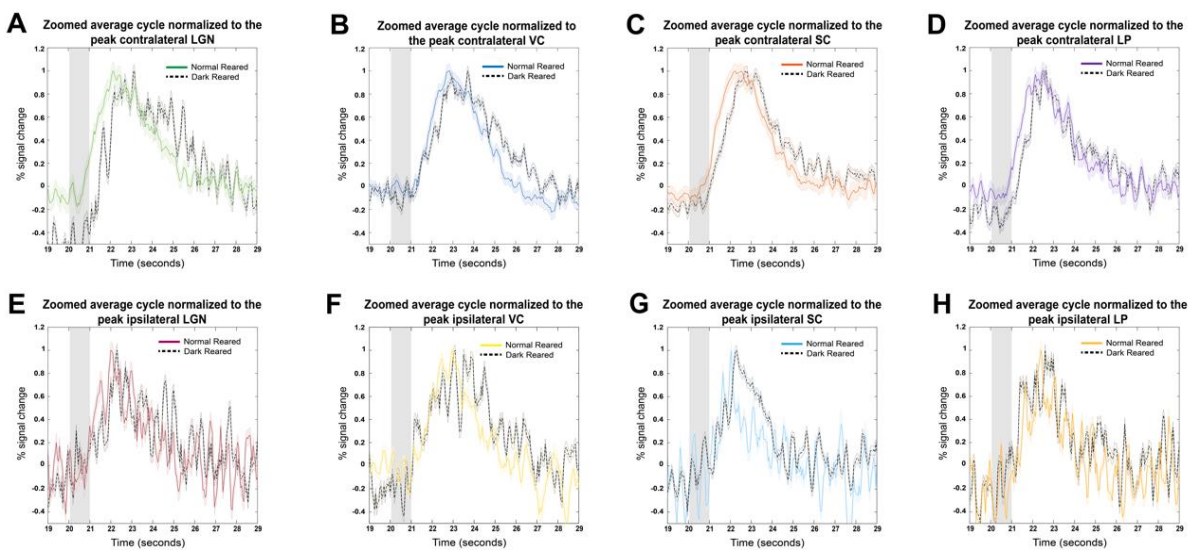

Fig.S6: Comparison of the averaged cycle (mean +/- s.e.m.) normalized to the maximum of individual ROIs with different rearing conditions: **(A)** contralateral LGN **(B)** contralateral VC **(C)** contralateral SC **(D)** contralateral LP **(E)** ipsilateral LGN **(F)** ipsilateral VC **(G)** ipsilateral SC **(H)** ipsilateral LP. An increased onset delay and overall response delay can be observed in most contralateral responses with minimal shift in the contralateral VC responses where only delayed peak is evident. Normalized ipsilateral responses for the two rearing conditions are similar in all ROIs.

31